\documentclass[a4paper,11pt]{article}
\usepackage{pos}
\usepackage{cleveref}
\usepackage{graphicx}
\usepackage{subfig}
\usepackage{booktabs}
\usepackage{makecell}

\title{Two-loop amplitude for the associated production of
a top-anti-top pair and a $W$ boson at hadron colliders}
\ShortTitle{Two-loop amplitude for $pp \to t\bar{t}W$}

\author*[a]{Mattia Pozzoli}

\affiliation[a]{Dipartimento di Fisica e Astronomia, Universit\`{a}	 di Bologna, \\
INFN, Sezione di Bologna, \\
via Irnerio 46, I-40126 Bologna, Italy}

\emailAdd{mattia.pozzoli@unibo.it}

\abstract{In this contribution I present a calculation of the inclusive cross-section for the associated production of a top-anti-top pair and a W boson at next-to-next-to-leading order in QCD, based on the first exact computation of the two-loop QCD amplitude at leading colour. I discuss a strategy to express the amplitude in terms of a set of special functions with rational coefficients, overcoming the obstacle posed by the presence of elliptic functions. The special functions are evaluated through the method of differential equations, while the rational coefficients are computed using finite field techniques. For the computation of the inclusive cross-section, the double-virtual correction is evaluated by interpolating a five-dimensional grid.}

\FullConference{The 33rd International Workshop on Deep Inelastic Scattering and Related Subjects (DIS2026)\\
4 - 8 May 2026\\
Bologna, Italy\\}

\begin{document}
\maketitle

\section{Introduction}
The production of a top-anti-top pair together with a $W$ boson ($t\bar{t}W$) is a process relevant both for searches for physics beyond the Standard Model~\cite{Buckley:2015lku,Dror:2015nkp,BessidskaiaBylund:2016jvp}, and as a background for the associated production of a top-anti-top pair with a Higgs boson ($t\bar{t}H$) and for four-top production ($t\bar{t}t\bar{t}$).

Moreover, there is tension between the experimental measurements of the inclusive cross-section~\cite{CMS:2022tkv,ATLAS:2024moy} and the NNLO QCD+NLO EW theoretical predictions~\cite{Buonocore:2023ljm}. In~\cite{Buonocore:2023ljm}, the two-loop QCD amplitude was computed by combining the soft-$W$ approximation~\cite{Catani:2022mfv,Barnreuther:2013qvf} (SA) and the procedure of massification~\cite{Penin:2005eh,Mitov:2006xs,Becher:2007cu} (MA). Within the corresponding systematic uncertainties, these dynamic approximations are expected to be adequate for the computation of the inclusive cross-section. Nevertheless, a calculation employing the exact two-loop amplitude is necessary for the validation of the approximations, and to compute differential distributions.

In this contribution, I present theoretical predictions for the inclusive cross-section at NNLO QCD~\cite{Becchetti:2026awn}, using the exact leading-colour (LC) two-loop QCD amplitude~\cite{Becchetti:2026yxl}. The computation is complicated by the presence of seven kinematic variables, which leads to intricate algebraic expressions, and of massive internal propagators in the Feynman diagrams, that yield loop integrals associated with elliptic curves. As was observed in the computation of the two-loop integrals appearing at LC~\cite{Becchetti:2025qlu}, there is an interplay between these sources of complexity, since the differential equations (DEs)~\cite{Barucchi:1973zm,Kotikov:1990kg,Kotikov:1991hm,Gehrmann:1999as,Bern:1993kr,Henn:2013pwa} satisfied by the master integrals (MIs) are plagued by gigantic expressions. As a consequence, the evaluation of the two-loop amplitude at a single phase-space point is significantly time-consuming. To compute the cross-section, we thus resort to an interpolation grid for the double-virtual correction.

\section{Setup and conventions}
We consider the scattering processes
\begin{equation}
\begin{split}
\bar{u}(p_1) + d(p_2) + \bar{t}(p_3)+t(p_4)+W^+(p_5) \longrightarrow 0,\\
\bar{d}(p_1) + u(p_2) + \bar{t}(p_3)+t(p_4)+W^-(p_5) \longrightarrow 0,
\end{split}
\label{eq:scattering_process}
\end{equation}
with all external momenta taken to be incoming, such that momentum conservation implies
\begin{equation}
p_1+p_2+p_3+p_4+p_5 = 0.
\label{eq:momentum conservation}
\end{equation}
The on-shell conditions read
\begin{equation}
p_1^2 = p_2^2 = 0, \quad p_3^2 = p_4^2 = m_t^2, \quad p_5^2 = m_W^2.
\label{eq:on_shell}
\end{equation}
In order to describe the kinematics of the process, we choose five Mandelstam variables $s_{ij} = (p_i +p_j)^2$ and the masses of the top quark and of the $W$ boson
\begin{equation}
    \vec{x} := 
    \left\{s_{13} , s_{34} , s_{24} , s_{25} , s_{15}, m_t^2 , m_W^2 \right\}.
\end{equation}
We treat the divergences of the loop integrals using dimensional regularisation, i.e.~we work in $D=4-2 \varepsilon$ spacetime dimensions.

We consider the generalised LC limit, i.e.~we expand the two-loop amplitude in the number of colours ($N_c=3$) and of light-quark flavours ($N_f=5$) as
\begin{equation}
\mathcal{A}^{(2)} = N_c^2 \mathcal{A}^{(2,N_c^2)}+N_c N_f \mathcal{A}^{(2,N_c N_f)}+N_f^2 \mathcal{A}^{(2,N_f^2)} + \mathcal{O}(N_c),
\label{eq:leading_colour}
\end{equation}
neglecting the terms suppressed by powers of $N_c$ and $N_f$. Unlike the SA and the MA, this parametric approximation is valid at every phase-space point, with the exception of threshold regions where sub-leading colour effects may be enhanced. While capturing the dominant behaviour of the amplitude, the LC approximation greatly simplifies the calculation: only 210 out of the 722 two-loop Feynman diagrams contribute at LC.

Working in the 't Hooft-Veltman (tHV) scheme~\cite{tHooft:1972tcz}, we use the method of physical projectors~\cite{Peraro:2019cjj,Peraro:2020sfm} to decompose the LC amplitude into 24 independent tensor structures $T_i$, chosen in accordance with the one-loop calculation of~\cite{Becchetti:2025osw}
\begin{equation}
\mathcal{A}^{(2)} = \sum_{i=1}^{24} F^{(i)} T_i.
\label{eq:ff_decomposition}
\end{equation}
The form factors $F^{(i)}$ are functions of the kinematics, independent of the polarisations of the external particles. These are expressed as linear combinations of 8959 scalar loop integrals as
\begin{equation}
F^{(i)} = \sum_{j=1}^{8959} r_j^{(i)}(\vec{x};\varepsilon) \ I_j(\vec{x};\varepsilon).
\label{eq:ff_FIs}
\end{equation}
After renormalisation and subtraction of the infrared (IR) singularities, we obtain the finite remainder
\begin{equation}
\mathcal{A}^{(2)}_\mathrm{fin} = \sum_{i=1}^{24} \sum_j q_j^{(i)}(\vec{x}) \ I_j^{(w_j)} (\vec{x}) \ T_i,
\label{eq:finrem}
\end{equation}
where $I_j^{(w_j)}$ is the coefficient of $\varepsilon^{w_j}$ in the Laurent expansion of the loop integral $I_j$, and $q_j^{(i)}$ is a rational function.

\section{Calculation of the amplitude}
Based on their set of inverse propagators, we group the loop integrals into integral families. Each element of an integral family is defined by specifying the integer exponents to which the inverse propagators are raised:
\begin{equation}
I_{a_1,\dots, a_s} = \int \left( \prod_{l=1}^L \frac{\mathrm{d}^{D}k_l e^{\varepsilon \gamma_E}}{\mathrm{i} \pi^{\frac{D}{2}}} \right) \frac{1}{\prod_{i=1}^s D_i^{a_i}}, \quad (a_1, \dots, a_s) \in \mathbb{Z}^s,
\label{eq:integral_family}
\end{equation}
where $L$ is the number of loops and $s$ is the number of inverse propagators $D_i$. The two-loop integrals either belong to the three irreducible two-loop families that were studied in~\cite{Becchetti:2025qlu}, whose inverse propagators are defined in \cref{tab:Intfam1}, or they can be written as products of two one-loop integrals of family ${\rm A}^{(\textrm{x12})}$, which were computed in~\cite{Becchetti:2025osw}. The set of inverse propagators for this one-loop family are defined in \cref{tab:Intfam1L}.

\begin{table}[t]
    \centering
    \begin{tabular}{c|c|c|c}
       & ${\rm F}_1$ & ${\rm F}_2$ & ${\rm F}_3$ \\
    \hline 
    $D_1$ & $k_1^2-m_t^2$ & $k_1^2$  & $k_1^2$ \\
    $D_2$ & $(k_1-p_3)^2$ & $(k_1-p_4)^2-m_t^2$  & $(k_1-p_2)^2$ \\
    $D_3$ & $(k_1-p_{23})^2$ & $(k_1-p_{34})^2$  & $(k_1-p_{25})^2$ \\
    $D_4$ & $(k_1-p_{235})^2$ & $(k_1-p_{234})^2$  & $(k_1+p_{34})^2$ \\
    $D_5$ & $k_2^2-m_t^2$ & $k_2^2$  & $k_2^2$ \\
    $D_6$ & $(k_2-p_4)^2$ & $(k_2+p_{2345})^2$  & $(k_2-p_3)^2-m_t^2$ \\
    $D_7$ & $(k_2+p_{235})^2$ & $(k_2+p_{234})^2$  & $(k_2-p_{34})^2$ \\
    $D_8$ & $(k_1+k_2)^2$ & $(k_1+k_2)^2$  & $(k_1+k_2)^2$ \\
    $D_9$ & $(k_1+p_4)^2-m_t^2$ & $(k_1-p_{2345})^2$  & $(k_1+p_3)^2-m_t^2$ \\
    $D_{10}$ & $(k_2+p_3)^2-m_t^2$ & $(k_2+p_4)^2-m_t^2$  & $(k_2+p_2)^2$ \\
    $D_{11}$ & $(k_2+p_{23})^2-m_t^2$ & $(k_2+p_{34})^2$  & $(k_2+p_{25})^2$ \\
\end{tabular}
    \caption{Inverse propagators of the two-loop irreducible integral families of~\cite{Becchetti:2025qlu}.
     We use the shorthand $p_{i_1\cdots i_k} = p_{i_1} + \dots + p_{i_k}$}
    \label{tab:Intfam1}
\end{table}

Not all loop integrals are linearly independent, as they satisfy linear relations, in particular integration-by-parts identities (IBPs)~\cite{Tkachov:1981wb,Chetyrkin:1981qh,Laporta:2001dd}. We exploit these relations to write the 8959 loop integrals in \cref{eq:ff_FIs} as linear combinations of 330 master integrals (MIs), which we denote by $\vec{\mathcal{I}}$. We perform this reduction to MIs using \textsc{NeatIBP}~\cite{Wu:2023upw} to generate the IBP system and \textsc{FiniteFlow}~\cite{Peraro:2019svx} to solve it.

\begin{table}[t]
    \centering
    \begin{tabular}{c|c}
      & ${\rm A}^{(\textrm{x12})}$  \\
    \hline 
    $D_1$ & $k_1^2$ \\
    $D_2$ & $(k_1-p_{1345})^2$ \\
    $D_3$ & $(k_1-p_{145})^2 -m_t^2$\\
    $D_4$ & $(k_1-p_{15})^2$\\
    $D_5$ & $(k_1-p_{5})^2$\\
\end{tabular}
    \caption{Inverse propagators of the one-loop irreducible integral family of~\cite{Becchetti:2025osw}. We use the shorthand $p_{i_1\cdots i_k}= p_{i_1} + \dots + p_{i_k}$}
    \label{tab:Intfam1L}
\end{table}

Furthermore, we exploit the IBPs to construct a system of linear differential equations (DEs)~\cite{Barucchi:1973zm,Kotikov:1990kg,Kotikov:1991hm,Gehrmann:1999as,Bern:1993kr,Henn:2013pwa} for the MIs
\begin{equation}
\forall \ \xi \in \vec{x} : \partial_{\xi} \vec{\mathcal{I}} (\vec{x};\varepsilon)=B_{\xi} (\vec{x};\varepsilon)\cdot  \vec{\mathcal{I}}(\vec{x};\varepsilon).
\label{eq:DEs}
\end{equation}
We then evaluate the MIs by solving the DEs. This task is simplified by choosing a basis of MIs satisfying DEs in canonical form~\cite{Henn:2013pwa}
\begin{equation}
\mathrm{d} \vec{\mathcal{I}}(\vec{x}; \varepsilon) = \varepsilon \ \mathrm{d} \tilde{A} (\vec{x}) \cdot \vec{\mathcal{I}} (\vec{x};\varepsilon),
\label{eq:canonical_basis}
\end{equation}
in which the dependence on the dimensional regulator $\varepsilon$ and on the kinematics $\vec{x}$ decouple, and the connection matrix $\tilde{A} (\vec{x})$ does not contain spurious poles. In the simplest case, which we refer to as polylogarithmic, the connection matrix takes the form
\begin{equation}
\mathrm{d} \tilde{A}(\vec{x}) = \sum_i a^{(i)} \ \mathrm{d} \log W_i (\vec{x}), \quad \mathrm{d} \log (z-c) = \frac{\mathrm{d} z}{z-c},
\label{eq:dlog_connection}
\end{equation}
where $a^{(i)}$ is a matrix of rational numbers, and $W_i (\vec{x})$ is an algebraic function.

The polylogarithmic case is well-understood, and for polylogarithmic integral families it is usually possible to construct a basis of MIs satisfying canonical DEs. However, the two-loop integral families of~\cite{Becchetti:2025qlu} involve also integrals associated with elliptic curves, which bring about differential forms such as
\begin{equation}
\frac{\mathrm{d}z}{\sqrt{\mathcal{P}_4(z)}}, \quad \mathrm{with} \quad \mathcal{P}_4(z) = (z-a_1) (z-a_2) (z-a_3) (z-a_4),
\label{eq:elliptic_curve}
\end{equation}
where the roots $a_1, \dots, a_4$ of the polynomial $\mathcal{P}_4(z)$ are non-degenerate. Canonical DEs for elliptic integral families have been obtained so far only in one case~\cite{Becchetti:2025oyb}. For this reason, the MIs of the two-loop integral families studied in~\cite{Becchetti:2025qlu} were chosen following the strategy of~\cite{Badger:2024fgb}, such that the elliptic MIs are finite and satisfy DEs quadratic in $\varepsilon$, while the polylogarithmic MIs obey canonical DEs.

This choice of MIs allows us to employ the method of~\cite{Badger:2024dxo} to express the loop integrals in terms of special functions. As a reminder, in order to compute the finite remainder of \cref{eq:finrem} we have to expand the MIs around $\varepsilon=0$ and subtract the ultraviolet (UV) and IR poles. Normalising the MIs such that the lowest order of the expansion is $\varepsilon^0$, in this case it suffices to truncate the expansion at $\varepsilon^4$
\begin{equation}
\mathcal{I}_j(\vec{x}; \varepsilon) =  \sum_{w=0}^{4}  \varepsilon^w \ \mathcal{I}_j^{(w)}(\vec{x}).
\label{eq:Laurent_expansion}
\end{equation}
The master integral coefficients $\mathcal{I}_j^{(w)}$ are not algebraically independent, introducing a redundancy in the representation of the MIs. For polylogarithmic MIs satisfying canonical DEs, it is possible to construct a basis of algebraically independent special functions $\{f_k^{(w)} \} \subset \{ I_j^{(w)} \}$ following the method of~\cite{Gehrmann:2018yef,Chicherin:2020oor,Chicherin:2021dyp,Abreu:2023rco}, such that any master integral coefficient $\mathcal{I}_j^{(w)}$ is expressed as a polynomial in the special functions and the zeta-values $\zeta_n$.

\begin{table}[t]
\begin{center}
\begin{tabular}{ccccccc}
\toprule
 & $f_k^{(1)}$ & $f_k^{(2)}$ & $f_k^{(3)}$ & $f_k^{(4)}$ & $f_k^{(4,*)}$ & Total\\
\midrule
\text{Form factors} & 7 & 12 & 63 & 212 & 29 & 323\\
\text{Finite remainder} & 7 & 12 & 63 & 187 & 29 & 298\\
\bottomrule
\end{tabular}
\caption{Number of special functions at each order appearing in the form factors and in the finite remainder.}\label{tab:special}
\end{center}
\end{table}

We extend the construction to our case following~\cite{Badger:2024dxo}. The choice of finite elliptic MIs allows us to carry out this construction for $w \leq 3$ in \cref{eq:Laurent_expansion}. At $w=4$, many integrals are polylogarithmic and we treat them analogously. The residual MIs, associated with elliptic curves, yield a few additional special functions that we label as $f_k^{(4,*)}$. This representation allows us to separate the polylogarithmic and the elliptic functions. Since the latter only appear in the finite part of the amplitude, and the former are algebraically independent by construction, it is possible to subtract the UV and IR poles analytically, obtaining the finite remainder of \cref{eq:finrem}. The number of special functions that appears in the form factors and in the finite remainder is reported in \cref{tab:special}.

We evaluate the special functions by solving the DEs they satisfy, which we derive from those obeyed by the MIs. Compared to the latter, the DEs for the special functions are faster to solve, since they are sparser and do not depend on $\varepsilon$. To solve these DEs, we compute the values of the MIs at some boundary point $\vec{x}_i$ using \textsc{AMFlow}~\cite{Liu:2022chg}. We then parametrise a linear path
\begin{equation}
\gamma(\eta) = (1- \eta) \vec{x}_i - \eta \vec{x}_f
\end{equation}
from the boundary point to a target point $\vec{x}_f$, yielding DEs in one variable along the path. We then use the DE solver of \textsc{AMFlow} to evolve the solution along the path, using the method of generalised series expansions~\cite{Pozzorini:2005ff,Moriello:2019yhu}. Because of the algebraic complexity of the DEs, the evaluation of the special functions is a bottleneck, as it requires around an hour per phase-space point.

At this stage, by performing IBP reduction and Laurent-expanding the MIs around $\varepsilon$ we express the form factors of \cref{eq:ff_FIs} and the finite remainder of \cref{eq:finrem} as polynomials in the special functions of \cref{tab:special} (and $\zeta_n$) whose coefficients are rational functions of $\vec{x}$. Obtaining the analytic expression of these coefficients requires us to manipulate large rational expressions in eight variables ($\vec{x}$ and $\varepsilon$). This is not feasible in this case, and we resort instead to the method described in~\cite{Peraro:2019okx}. At each phase-space point, for rationalised values of the kinematic invariants $\vec{x}$, we reconstruct the exact numerical value of the coefficients from numerical evaluations modulo prime numbers~\cite{vonManteuffel:2014ixa,Peraro:2016wsq} using \textsc{FiniteFlow}~\cite{Peraro:2019svx}. The number of prime numbers required is sensitive to the rationalisation precision of the invariants and to the degree of the coefficients. Due to the large number of primes needed (up to 400 for a single phase-space point), the evaluation time of the coefficients is around 10 minutes per phase-space point, which is however not a bottleneck compared to the time required to evaluate the special functions.

\section{From the amplitude to the cross-section}
In this section, we present results for the NNLO QCD inclusive cross-section at a centre-of-mass energy of \mbox{$\sqrt{s}=13$\,TeV}. Since the timings discussed in the previous section do not allow for an on-the-fly evaluation of the two-loop amplitude, for the computation of the inclusive cross-section we resort to an interpolation grid. We set the two masses to their physical values \mbox{$m_t = 173.2$\,GeV} and \mbox{$m_W = 80.385$\,GeV}, and we parametrise the phase-space in terms of two energy fractions and three angles according to~\cite{Agarwal:2024jyq}. We then compute the finite remainder at 224'640 phase-space points, yielding a five-dimensional grid in those variables. We validate the interpolation at NLO against {\sc OpenLoops}~\cite{Buccioni:2019sur}, and we estimate the uncertainty to be below the percent level.

\renewcommand{\arraystretch}{1.5}
\begin{table}[t]
\centering
\begin{tabular}{|c|c|c|}
\hline
        $\sigma_{\rm NNLO \ QCD}$[fb] & $t{\bar t}W^-$ & $t{\bar t}W^+$ \\
\hline
	SA+MA & $235.4(0)^{+5.1\%}_{-6.6\%}\pm 1.9\%$ & $474.9(2)^{+4.8\%}_{-6.4\%}\pm 1.9\%$ \\
\hline
	LC & $241.9(0)^{+6.4\%}_{-7.3\%}\pm 2.3\%$ & $489.4(1)^{+6.3\%}_{-7.2\%}\pm 2.5\%$ \\
\hline
\end{tabular}
\caption{Results for the NNLO QCD inclusive cross-section based on the exact LC two-loop amplitude (second row), compared with those of~\cite{Buonocore:2023ljm} (first row), which are computed combining the SA and the MA.}
\label{tab:xs}
\end{table}

In order to compare our results with~\cite{Buonocore:2023ljm}, we work with the same setup, with the exception of the double-virtual correction, which we evaluate using the interpolation grid. We compute the cross-section within the {\sc Matrix} framework \cite{Grazzini:2017mhc}, employing the $q_T$-subtraction formalism~\cite{Catani:2007vq} to subtract the IR singularities. The results for the inclusive cross-section are shown in \cref{tab:xs}. Within the uncertainties, the computation based on the exact two-loop LC amplitude is compatible with the predictions of~\cite{Buonocore:2023ljm}, based on the combination of the SA and the MA. 

We estimate the perturbative uncertainties through seven-point scale variation. Additionally, at NLO the comparison between the LC and the exact result indicates that sub-leading colour effects amount to around 22\% of the virtual correction, and that they are uniform across phase-space. We observe an analogous behaviour by comparing distributions based on the LC result at one and two loops, hinting to the fact that no new distortions arise at two loops. Therefore, we estimate the impact of the missing sub-leading colour effects on the double-virtual correction to be 22\% of the LC result. As the size of the two-loop virtual correction is around 10\% of the NNLO QCD cross-section, the uncertainty associated with the missing sub-leading colour contributions is roughly 2.2\%.

\section{Conclusion}
In this contribution, I presented a framework for the numerical evaluation of the leading-colour two-loop QCD amplitude for $t\bar{t}W$ hadroproduction~\cite{Becchetti:2026yxl}. We expressed the amplitude in terms of a set of special functions, evaluated through generalised series expansion, with rational coefficients, whose exact value is reconstructed point-by-point from finite-field evaluations. The time required to evaluate the finite remainder at a phase-space point is of the order of one hour. Furthermore, we produced an interpolation grid for the finite remainder, which we used to compute the inclusive cross-section at NNLO QCD~\cite{Becchetti:2026awn}. The results are in agreement with those of~\cite{Buonocore:2023ljm}.

\acknowledgments

I am grateful to my collaborators M. Becchetti, D. Canko, X. Chen, V. Chestnov, M. Delto, S. Ditsch, M. Grazzini, S. Kallweit, T. Peraro, C. Savoini, L. Tancredi and S. Zoia for their contribution to the results presented in these proceedings. This work was supported by the European Research Council (ERC) under the European Union's Horizon Europe research and innovation program grant agreement 101040760, \textit{High-precision multi-leg Higgs and top physics with finite fields} (ERC Starting Grant \emph{FFHiggsTop}).

\bibliographystyle{JHEP}
\bibliography{biblio}

\end{document}